\documentclass[12pt]{iopart}

\usepackage{iopams}
\newtheorem{lemma}{Lemma}
\newtheorem{proposition}{Proposition}

\newcommand{\be}{\begin{equation}}
\newcommand{\ee}{\end{equation}}
\def\tr{\mathop{\rm tr}\nolimits}

\def\dif{{\rm d}}

\usepackage[latin1]{inputenc}

\begin{document}

\title[Dimension of the isometry group in 3-D Riemannian spaces]
{Dimension of the isometry group in three-dimensional Riemannian spaces}

\author{Joan Josep Ferrando$^{1,2}$ and  Juan Antonio S\'aez$^3$}

\address{$^1$\ Departament d'Astronomia i Astrof\'{\i}sica, Universitat
de Val\`encia, E-46100 Burjassot, Val\`encia, Spain}

\address{$^2$\ Observatori Astron\`omic, Universitat
de Val\`encia, E-46980 Paterna, Val\`encia, Spain}

\address{$^3$\ Departament de Matem\`atiques per a l'Economia i l'Empresa,
Universitat de Val\`encia, E-46022 Val\`encia, Spain}

\ead{joan.ferrando@uv.es; juan.a.saez@uv.es}

\begin{abstract}
The necessary and sufficient conditions for a three-dimensional Riemannian metric to admit a group of isometries of dimension $r$ acting on s-dimensional orbits are obtained. These conditions are Intrinsic, Deductive, Explicit and ALgorithmic and they offer an IDEAL labeling that improves previously known invariant studies.
\end{abstract}
%
%


%


\section{Introduction}
\label{sec-intro}
The invariant characterization of the three-dimensional Riemannian metrics admitting a group G$_r$ of isometries acting on orbits O$_s$ was presented by Bona and Coll years ago \cite{bonacoll1, bonacoll2}. Their approach is supported by the capital theorems by Eisenhart \cite{eisenhart} and Kerr \cite{Kerr}, and they employ conditions that are expressed in terms of the eigenvalues and eigenvectors of the Ricci tensor. This kind of invariant approach has been revisited recently \cite{Tomoda} by offering some algorithms for computing the dimension of the isometry group. Lately \cite{FS-G3} we have presented an IDEAL approach to the transitive case: we have given the necessary and sufficient (Intrinsic, Deductive, Explicit and ALgorithmic) conditions for a three-dimensional Riemannian metric to admit a transitive group of isometries, and we have also distinguished the three different groups G$_6$, the three different groups G$_4$ and the ten Bianchi-Behr types G$_3$ in transitive action. Here we extend our IDEAL labeling to the non-transitive case by distinguishing the action of a G$_3$ on two-dimensional orbits, the existence of a group G$_2$ and the action of a group G$_1$. We summarize our results in a compact algorithm that uses explicit metric tensorial concomitants of the Ricci tensor. We also present explicit Ricci concomitants that allow us to distinguish between the three cases of G$_3$ on O$_2$ and to discriminate when the group G$_2$ is abelian. Our tensorial approach avoids obtaining the Ricci eigenvectors and eigenvalues, which are necessary in the algorithms presented in the recent paper by Kruglikov and Tomoda \cite{Tomoda}. 

We name Ricci-frame an orthonormal frame that can be obtained from the Ricci tensor and its covariant derivatives. The Ricci-frames play an important role in studying the dimension of the isometry group. Indeed, the results by Bona and Coll \cite{bonacoll1, bonacoll2} state that, when the action is simply transitive on the orbits, this dimension depends on the number of independent functions generated by the connection coefficients (and their first and second derivatives) defined by a Ricci-frame. In section \ref{sec-Z} we associate to any orthonormal frame a connection tensor $Z$ and two of its differential concomitants that collect, respectively, the connection coefficient of the frame and their first and second directional derivatives. These tensors allow us to characterize the number of independent functions generated by the connection coefficients.

The different cases in which a Ricci-frame exists are analyzed in section \ref{sec-obtaining} and, for each of them, we obtain an explicit concomitant $T$ of the Ricci tensor with this Ricci-frame as eigenframe. From a previous result \cite{FS-G3}, we can determine the connection tensor $Z$ in terms of this tensor $T$, and consequently, we obtain $Z$ as a concomitant of the Ricci tensor.

With the results obtained in sections \ref{sec-Z} and \ref{sec-obtaining} we can perform an algorithm, presented in section \ref{sec-diagram} as a flow diagram, that distinguishes the three-dimensional Riemannian spaces admitting a group of isometries G$_r$ acting on s-dimensional orbits. Finally, section \ref{sec-G3(2)-G2} is devoted to expressing in terms of explicit Ricci concomitants the conditions distinguishing when an isometry group G$_2$ is commutative, and the sign of the curvature of the orbits of a G$_3$ acting on O$_2$.


\section{The connection tensor and its differential concomitants}
\label{sec-Z}
In an oriented three-dimensional Riemannian manifold with metric $g$ and volume element $\eta$ let us
consider $\{ e_a \}$, an oriented ($\eta=e_1 \wedge e_2 \wedge e_3$)
orthonormal frame of vector fields, and $\{ \theta^a \}$, to be its dual
basis. The connection coefficients $\gamma^{c}_{ab}$ are defined as
usual by
\begin{equation} \label{c-c}
\nabla e_a = \gamma^{c}_{ab} \  \theta^b \otimes \e_c \, .
\end{equation}
Associated with the frame $\{ e_a \}$ we can define its 
{\it connection tensor} $Z$ as
\begin{equation} \label{zeta}
\hspace{-1cm} Z \equiv  \frac{1}{2} \epsilon^{abc} \Big[( \nabla e_a ) \cdot e_b
\Big] \otimes e_c \, , \quad Z={Z_a}^b \, \theta^a \otimes e_b\, , \quad
{Z_a}^b=\frac{1}{2} \,\epsilon^{cdb} \,  \delta_{de} 
\, \gamma^{e}_{ca} \, ,
\end{equation}
where $\epsilon^{abc}$ is the Levi-Civita symbol. In what follows, a $\cdot$ denotes
the contraction of the adjacent indexes in the tensorial product, and
$A^2 = A \cdot A$. Tensor $Z$ is invariant when we change the orthonormal frame with a constant rotation. Moreover, it holds that
\begin{equation} \label{derivada}
\nabla_{i} (e_a)_j = (e_a)^k \, {Z_{i}}^{l} \eta_{l k j} \, .
\end{equation}
Note that indexes $a,b,\dots$ are used to count the vectors of
the frame, and indexes $i, j , \dots$ will indicate components in a
coordinate frame. We shall denote with the same symbol a tensor and
its associated tensors by raising and lowering indexes with the
metric $g$.  

Tensor $Z$ has been considered previously in \cite{FS-G3} and, if a transitive group G$_3$ leaving the frame invariant exists, it coincides with the {\it structure tensor} of G$_3$.

The last expression in (\ref{zeta}) implies that each of the $3^2$ components ${Z_a}^b$ of the connection tensor $Z$ in the frame $\{ e_a \}$ corresponds with each of the nine connection coefficients $\gamma^{c}_{ab}$ of this frame. Now we define two tensorial differential concomitants of $Z$, $C(Z)$ and $D(Z)$, which collect, respectively, the $3^3$ first derivatives, $e{_a}({Z_b}^c)$, and the $3^4$ second derivatives, $e_d e{_a}({Z_b}^c)$, of the connection coefficients. More precisely, from the definition (\ref{zeta}) and expression (\ref{derivada}) we obtain:
\begin{proposition} \label{CD}
Let $Z$ be the connection tensor of the frame $\{e_a \}$
and let us define its differential concomitants
\begin{eqnarray} 
\hspace{-1.5cm} C = C(Z)  ,  \qquad \  C_{kij} \equiv
\nabla_k Z_{ij} + {Z_{k}}^m \, \Big( \eta_{mi}^{\ \ n} \, Z_{nj}+ \eta_{mj}^{\ \ n} Z_{in} \Big) \, , \label{C(Z)} \\
\hspace{-1.5cm}  D = D(Z)  ,  \qquad  D_{ijkl} \equiv \nabla_{i} C_{jkl} +
{Z_{i}}^{n} \Big( {\eta_{nj}}^m \, C_{mkl} + {\eta_{n k }}^m \, C_{j
m l} + {\eta_{n l}}^m \, C_{j k m   } \Big) \, . \label{D(Z)}
\end{eqnarray}
Then we have
\begin{equation}
C =  e_a( {Z_b}^c ) \ \theta^a \otimes \theta^b \otimes e_c \, , \qquad  D =  e_a  e_b ({Z_e}^g) \, \theta^a \otimes \theta^b  \otimes \theta^e \otimes e_g \, .
\end{equation}
\end{proposition}

The isotropy group is trivial if, and only if, a Ricci-frame exists. In this case, the dimension of the isometry group coincides with that of the orbits and it depends on the number of independent functions generated by the connection coefficients of the Ricci-frame and its first and second derivatives \cite{bonacoll1, bonacoll2}. Now we introduce some algebraic concomitants of tensors $C(Z)$ and $D(Z)$ that allow us to characterize this number. Indeed, the number of independent functions appearing in the connection coefficients and their derivatives is the number of directions generated by the first tensorial index of $C$ and $D$. This way, we can make them depend on one, two or three functions by simply imposing the nullity or non-nullity of the tensorial concomitant built as the contraction of the first index of each factor, of a linear, quadratic or cubic expression, with the volume element $\eta$. More precisely, we have:
\begin{proposition} \label{JKLMN}
Let $Z$ be the connection tensor of the frame $\{e_a \}$ and $C=C(Z)$, $D=D(Z)$ the tensors given in {\em (\ref{C(Z)})} and {\em (\ref{D(Z)})}. Let us define the concomitants
\begin{equation} 
%
{I(Z)_{ijkl}}^r= \, C_{pij } \,
C_{qkl}\,  \eta^{pqr},  \qquad \qquad \  {J(Z)_{ijklm}}^r= \, C_{pij } \, D_{qklm}\,  \eta^{pqr} ,   
\label{IJ}
\end{equation}
\begin{equation}
L(Z) = I(Z) \cdot C  \, ,  \qquad  M(Z) = I(Z) \cdot D \, , \qquad  
N(Z) = J(Z) \cdot D\, . \label{LMN} 
\end{equation}
Then, it holds
\begin{enumerate}
\item All the connection coefficients are constant if, and only if, $C(Z)=0$.
\item All the connection coefficients depend on a single function
$x$, $\gamma^{a}_{bc} = \gamma^{a}_{bc}(x)$, and $\dif x = X_a (x)
\theta^a$ if, and only if,
\be
 C(Z) \neq 0 \, , \qquad I(Z) =0 \, , \qquad J(Z) =0 \, .   \label{H2}
\ee
\item All the connection coefficients depend on two functions
$x, y$, $\gamma^{a}_{bc} = \gamma^{a}_{bc}(x,y)$ with $\dif x \wedge
\dif y \neq 0$, and  $\dif x =X_a (x,y) \theta^a$, $\dif y = Y_a
(x,y) \theta^a$ if, and only if, one of the two following conditions hold:
\be
\hspace{-2.5cm}
\{I(Z) \neq 0 , \   L(Z) =0 , \  M(Z) =0\}   \quad  {\rm or} \quad 
\{I(Z) = 0 , \  J(Z) \neq 0 , \  N(Z) =0 \}  \, .  \label{H1}
\ee
\item Otherwise, the connection coefficients define three
independent functions,
\end{enumerate}
\end{proposition}
%


\section{Obtaining the connection tensor $Z$}
\label{sec-obtaining}

In order to perform our IDEAL approach when the isotropy group is trivial, we must obtain the connection tensor $Z$ associated with a Ricci-frame as an explicit concomitant of the Ricci tensor. A cornerstone to achieve this is the following result proved in \cite{FS-G3}:
\begin{lemma} \label{lemma-T-zeta}
If $T$ is an algebraic general traceless tensor that diagonalizes in
the orthonormal frame $\{ e_a \}$, the connection tensor $Z$ corresponding to this frame
can be obtained as
\begin{equation} \label{Z(T)}
Z= Z(T) \equiv \frac{1}{e} K \! \cdot \! \Big[3 b \, T^2 + 6 c \, T + \frac12 b^2 \, g \Big]  \, . 
 \end{equation}
where
\begin{equation} \label{Z(T)-b}
K_{i}^{\ j} \equiv  (\nabla  T \cdot T)_{i k l} \eta^{l k j}  , \quad b \equiv \tr T^2 \, , \quad c \equiv \tr T^3 \,  , \quad e \equiv   b^3 - 6 c^2  \, .
\end{equation}
\end{lemma}

When tensor $T$ of the lemma above is a Ricci concomitant, then the orthonormal frame defined by its eigenvectors is a Ricci-frame. Now we analyze the different cases in which this tensor $T$ exists and how it can be obtained from the Ricci tensor.

Let $R$ be the Ricci tensor and let us consider the following Ricci concomitants:
\begin{equation} 
r\equiv \tr R \, , \qquad S\equiv   R- \frac{r}{3} g \, , \qquad  s \equiv \tr
S^2 \, , \qquad t = \tr S^3 \, . \label{Srst}
\end{equation}
If the Ricci tensor is algebraically general, $6 t^2 \not=s^3$, then its traceless part $S$ defines a Ricci-frame, and we can take  $T= S$ to compute $Z$, as considered in \cite{FS-G3}. 

In the algebraic special case, $6 t^2 =s^3$, we know \cite{bonacoll1, bonacoll2, FS-G3} that a group G$_6$ exists if the Ricci tensor is proportional to $g$ ($s=0$). And, when the Ricci tensor admits two different eigenvalues ($s \not=0$) and the simple eigenvector $u$ is not shear-free, a Ricci-frame can be found. Then, tensor $T$ in lemma \ref{lemma-T-zeta} can be taken as $T=\Sigma$, where \cite{FS-G3}:
\begin{equation} \label{Dsigma}
\label{Sigma}
\hspace{-2cm} \Sigma \equiv D - \frac12 (\tr D) \, h \, , \quad D_{ij} \equiv (\nabla R \cdot R)^{k l m} \eta_{m l (i}\, h_{j)k}   \, , \quad h \equiv \frac{1}{t} \Big(2 \, S^3 - \frac43 s \, S \Big) \, .
\end{equation}
The last expression above gives the projector on the Ricci eigenplane, $h = g - u \otimes u$, in terms of Ricci concomitants.

If the Ricci tensor is algebraically special and the simple eigenvector $u$ is shear-free, then a group G$_3$ acting simply transitively on the whole space does not exist \cite{bonacoll1, bonacoll2, FS-G3}. Nevertheless, groups G$_1$ and G$_2$ in action simply transitive on the orbits can exist in this case. Then, the corresponding Ricci-frames can be obtained from the gradient of the invariant scalars as we will see next.

Note that the Bianchi identities, $2 \, \nabla \cdot R = \dif r$, imply that the simple eigenvector $u$ is geodesic, $\dot{u}=0$, if, and only if, $\dif \alpha \wedge  u = 0$, where $\alpha$ is the simple Ricci eigenvalue. This invariant depends on the scalars $r$, $s$ and $t$ as:
\begin{equation} \label{alpha}
\alpha= \frac{1}{3} r + 2 \frac{t}{s} \, .
\end{equation}

Then, if the Ricci tensor is algebraically special and the simple eigenvector $u$ is shear-free and non-geodesic ($\dot{u} \neq 0$), a Ricci-frame exists. It is the one determined by $u$ and $h(\dif \alpha)$. In this case, tensor $T$ in lemma \ref{lemma-T-zeta} can be taken as $T=A$, where:
\begin{equation} \label{A}
 A=  2 \, h(\dif \alpha) \otimes h (\dif \alpha) - h(\dif \alpha, \dif \alpha) h \, , \qquad \dif \alpha = \frac{1}{3} \Big( \dif r + \frac{2}{s} \, \dif t \Big) \, .
\end{equation}
The expression for $\dif \alpha$ above can be obtained from (\ref{alpha}) by using that $6 t^2 = s^3$. 

Finally, let us suppose that $\dot{u} =0$. Then, we have that $(s\, \dif r + 2\, \dif t) \wedge u=0$, and the algebraically special condition, $6 t^2 = s^3$, implies that all the algebraic scalars associated with the Ricci tensor have a gradient in the direction
of $u$ if, and only if, $\dif r \wedge u =0$. If this is not the case, namely if $\dif r \wedge u \not=0$, a Ricci-frame exists. It is the one determined by $u$ and $h(\dif r)$. In this case, tensor $T$ in lemma \ref{lemma-T-zeta} can be taken as $T=Y$, where:
\begin{equation} \label{Y}
 Y=  2 \, h(\dif r) \otimes h (\dif r) - h(\dif r, \dif r) h \,  .
\end{equation}

If none of the above considered cases happens, that is, if the Ricci tensor is algebraically special, if the simple eigenvector $u$ is geodesic and shear-free and if $\dif r \wedge u =0$, then no Ricci-frame can be built \cite{bonacoll1, bonacoll2}. Then, two possibilities may occur. If $\dif r = \dif s =0$, we have that a G$_4$ exists \cite{bonacoll1, bonacoll2, FS-G3}. And, if $(\dif r)^2 + (\dif s)^2 \neq 0$, we have that a G$_3^{(2)}$ (three-dimensional group acting on two dimensional orbits) exists \cite{bonacoll1, bonacoll2}.

It is worth remarking that the above analysis on the existence of Ricci-frames was indicated by Bona and Coll \cite{bonacoll1, bonacoll2}. Here, we have delved into how they can be obtained and how to determine tensor $T$, which provides its associated connection tensor $Z$ (see lemma \ref{lemma-T-zeta}). Once this tensor has been obtained, the results in proposition \ref{JKLMN} provide an IDEAL labeling that we present in algorithmic form below.


\section{Algorithm to determine the dimension of the isometry group} 
\label{sec-diagram}

The IDEAL characterization of the Riemannian spaces with a non-trivial isotropy group is given in the section above: a G$_6$ when $3R = rg$, and a G$_4$ (respectively, a G$_3^{(2)}$) if $3R \not= rg$ and if there is no Ricci-frame and $\dif r = \dif s =0$ (respectively, $(\dif r)^2 + (\dif s)^2 \neq 0$). 

We have also considered the different cases in which a Ricci-frame exists. For each case, we have given the explicit expression (in terms of Ricci concomitants) of a trace-less algebraically general tensor $T$ that provides the connection tensor $Z$ associated with the Ricci-frame (lemma \ref{lemma-T-zeta}). Thus, in order to obtain the IDEAL characterization of the Riemannian spaces with a trivial isotropy group, and as consequence of the results by Bona and Coll \cite{bonacoll1, bonacoll2}, we must impose on $Z$ the constraints $C(Z)=0$, (\ref{H2}) or (\ref{H1}), which denote that the isometry group has dimension three, two or one, respectively. Note that, if the first condition does not hold ($C(Z) \not=0$), then (\ref{H2}) states:
\be
\hspace{-1.8cm} {\rm H2}  \qquad \qquad I(Z) =0 \, , \qquad J(Z) =0 \, .   \label{H2b}
\ee
And if the above condition H2 does not hold, then (\ref{H1}) states:
\be
\hspace{-1.8cm}  {\rm H1}  \qquad \qquad  \{I(Z) \not= 0 , \ L(Z) =0 , \  M(Z) =0\}   \ \  {\rm or} \ \ 
\{J(Z) \neq 0 , \  N(Z) =0 \}   \, .   \label{H1b}
\ee

With all these result we can build a flowchart that performs an algorithm providing the dimension of the isometry group of a three-dimensional Riemannian metric (figure 1). Note that all the involved conditions are given in terms of explicit Ricci concomitants, and consequently, this algorithm offers an IDEAL labeling of each geometry.


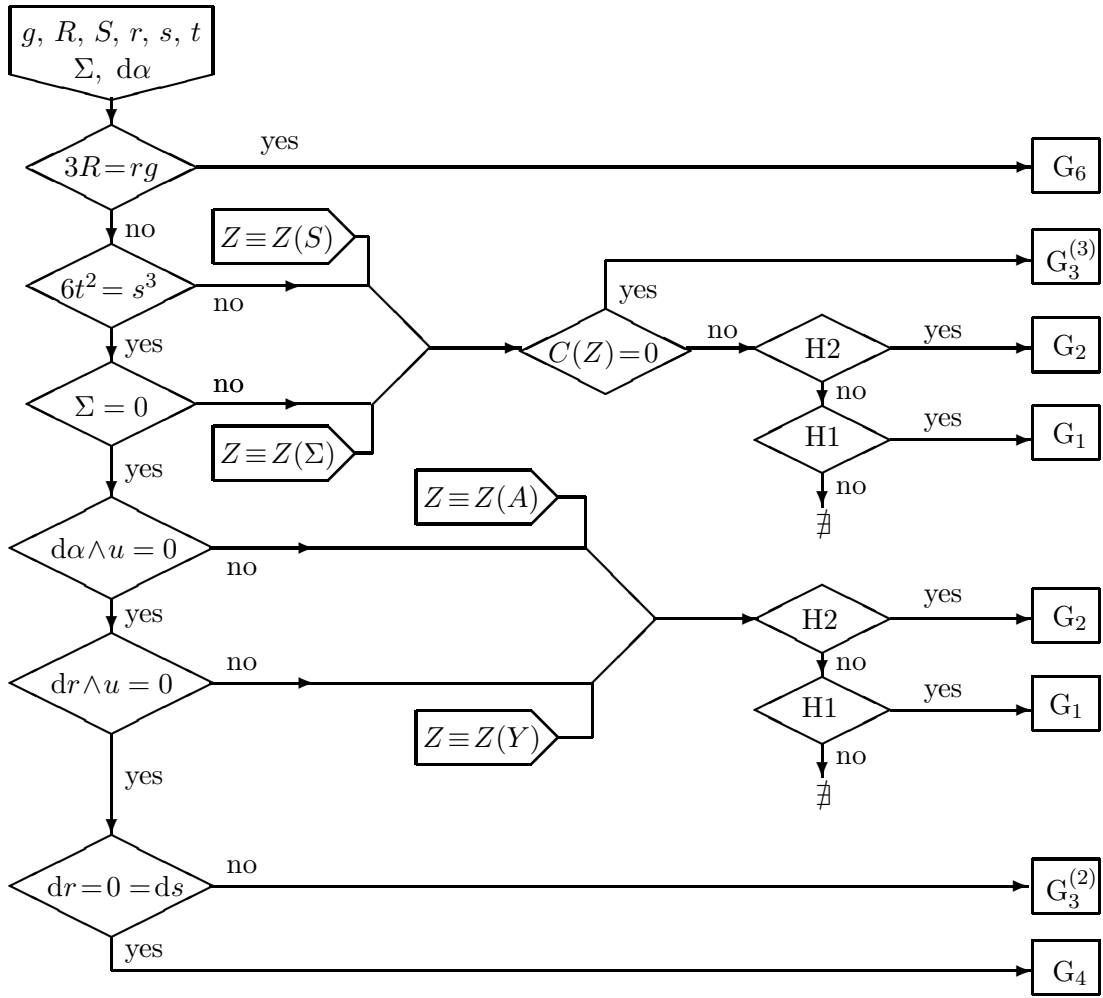
\begin{figure}[b]

\vspace*{0mm}

\hspace*{2mm} \setlength{\unitlength}{0.9cm} {\small \noindent
\begin{picture}(0,18)
\thicklines

\put(2.5,17){\line(-4,-1){1.5}}
 \put(-0.5,17){\line(4,-1){1.5}}
\put(-0.5,17){\line(0,1){1}} \put(2.5,18){\line(-1,0){3}}
\put(2.5,18){\line(0,-1){1}} \put(-0.4,17.5 ){$\, g , \, R  , \, S , \, r , \, s  , \,   t $}
 \put(0.15,16.95 ){$ \ \  \Sigma, \ \dif \alpha   $}
\put(1,16.65){\vector(0,-1){0.4}}

\put(1,16.25){\line(-2,-1){1.25}} \put(1,16.25){\line(2,-1){1.25}}
\put(1,15){\line(2,1){1.25}} \put(1,15){\line(-2,1){1.25}}
\put(0.3,15.5){$3 R \!  =\! r g $}


\put(1,14.5){\line(-2,-1){1.25}} \put(1,14.5){\line(2,-1){1.25}}
\put(1,13.25 ){\line(2,1){1.25}} \put(1,13.25 ){\line(-2,1){1.25}}
\put(0.25,13.7){$6 t^2 \! = s^3 $}

\put(1,15){\vector(0,-1){0.5}}
 \put(1,13.25 ){\vector(0,-1){0.5}}


\put(1,12.75){\line(-2,-1){1.25}} \put(1,12.75){\line(2,-1){1.25}}
\put(1,11.5 ){\line(2,1){1.25}} \put(1,11.5 ){\line(-2,1){1.25}}
\put(0.45,11.95){$\Sigma = 0 $}

\put(11.5,12.45){\vector(0,-1){0.35}}


\put(11.4,10.2){$\nexists$}

\put(11.5,11.1){\vector(0,-1){0.5}}

\put(8.3,13.53 ){\line(-2,-1){1.25}} \put(8.3,13.53
){\line(2,-1){1.25}} \put(8.3,12.28){\line(2,1){1.25}}
\put(8.3,12.28 ){\line(-2,1){1.27}} \put(7.45 ,12.75){$C(Z)\!=\!0$}



\put(11.5,13.45 ){\line(-2,-1){1}} \put(11.5,13.45 ){\line(2,-1){1}}
\put(11.5,12.45){\line(2,1){1}} \put(11.5,12.45 ){\line(-2,1){1}}
\put(11.1 ,12.8){ H2 }

\put(11.5,12.1 ){\line(-2,-1){1}} \put(11.5,12.1 ){\line(2,-1){1}}
\put(11.5,11.1){\line(2,1){1}} \put(11.5,11.1){\line(-2,1){1}}
\put(11.1 ,11.5){ H1 }

\put(2.25,15.63){\vector(1,0){12.35}}

\put(2.25,13.88){\vector(1,0){1.56}}\put(3.81,13.88){\line(1,0){1}}

\put(2.25,12.13){\vector(1,0){1.56}}\put(3.81,12.13){\line(1,0){1.02}}

\put(4.8,13.88){\line(1,-1){0.92}}
\put(4.85,12.11){\line(1,1){0.85}}
\put(5.7,12.95 ){\vector(1,0){1.35}}


\put(9.5,12.95 ){\vector(1,0){1}}

\put(12.5,12.95 ){\vector(1,0){2.1}}
 \put(12.5,11.6){\vector(1,0){2.1}}

\put(8.3,14.25){\line(0,-1){0.7}} \put(8.3,14.25){\vector(1,0){6.3}}

\put(14.6,15.25){\line(1,0){1}} \put(14.6,15.25 ){\line(0,1){0.8}}
\put(15.6,16.05 ){\line(-1,0){1}} \put(15.6,16.05
){\line(0,-1){0.8}} \put(14.9,15.5){G$_6$}

\put(14.6,13.9){\line(1,0){1}} \put(14.6,13.9){\line(0,1){0.8}}
\put(15.6,14.7 ){\line(-1,0){1}} \put(15.6,14.7){\line(0,-1){0.8}}
\put(14.8,14.1){G$_3^{(3)}$}

\put(14.6,12.6){\line(1,0){1}} \put(14.6,12.6 ){\line(0,1){0.8}}
\put(15.6,13.4){\line(-1,0){1}} \put(15.6,13.4 ){\line(0,-1){0.8}}
\put(14.9,12.85){G$_2$}

\put(14.6,11.3){\line(1,0){1}} \put(14.6,11.3 ){\line(0,1){0.8}}
\put(15.6,12.1){\line(-1,0){1}} \put(15.6,12.1 ){\line(0,-1){0.8}}
\put(14.9,11.55){G$_1$}

\put(1,10.75){\line(-2,-1){1.5}} \put(1,10.75){\line(2,-1){1.5}}
\put(1,9.25 ){\line(2,1){1.5}} \put(1,9.25 ){\line(-2,1){1.5}}
\put(0.1,9.85){$\dif \alpha \! \wedge \! u  = 0 $}

\put(1,8.75){\line(-2,-1){1.5}} \put(1,8.75){\line(2,-1){1.5}}
\put(1,7.25 ){\line(2,1){1.5}} \put(1,7.25 ){\line(-2,1){1.5}}
\put(0.1,7.85){$\dif  r \! \wedge\! u  = 0 $}

\put(9,8.95 ){\vector(1,0){1.55}}


\put(2.45,10){\vector(1,0){1.55}}\put(4,10){\line(1,0){4}}

\put(2.45,8){\vector(1,0){1,55}}\put(4,8){\line(1,0){4.1}}

\put(8,10){\line(1,-1){1.05}}
 \put(8.1,8){\line(1,1){0.95}}




\put(11.5,9.45 ){\line(-2,-1){1}} \put(11.5,9.45 ){\line(2,-1){1}}
\put(11.5,8.45){\line(2,1){1}} \put(11.5,8.45 ){\line(-2,1){1}}
\put(11.2 ,8.8){H2}

\put(11.5,8.1 ){\line(-2,-1){1}} \put(11.5,8.1 ){\line(2,-1){1}}
\put(11.5,7.1){\line(2,1){1}} \put(11.5,7.1){\line(-2,1){1}}
\put(11.2 ,7.5){H1}

\put(14.6,8.6){\line(1,0){1}} \put(14.6,8.6 ){\line(0,1){0.8}}
\put(15.6,9.4){\line(-1,0){1}} \put(15.6,9.4 ){\line(0,-1){0.8}}
\put(14.9,8.85){G$_2$}

\put(14.6,7.3){\line(1,0){1}} \put(14.6,7.3 ){\line(0,1){0.8}}
\put(15.6,8.1){\line(-1,0){1}} \put(15.6,8.1 ){\line(0,-1){0.8}}
\put(14.85,7.55){G$_1$}

 \put(12.5,8.95){\vector(1,0){2.1}}
\put(12.5,7.6){\vector(1,0){2.1}}
\put(11.5,8.45){\vector(0,-1){0.35}}
\put(11.5,7.1){\vector(0,-1){0.5}} \put(11.4,6.2){$\nexists$}

\put(1,11.5){\vector(0,-1){0.75}}
 \put(1,9.25 ){\vector(0,-1){0.5}}


\put(1,5.75){\line(-2,-1){1.5}} \put(1,5.75){\line(2,-1){1.5}}
\put(1,4.25 ){\line(2,1){1.5}} \put(1,4.25 ){\line(-2,1){1.5}}
\put(0.05,4.85){$\dif r\!= \!0=\! \dif s $}


\put(1,7.25){\vector(0,-1){1.5}}


\put(14.6,4.6){\line(1,0){1}} \put(14.6,4.6 ){\line(0,1){0.8}}
\put(15.6,5.4){\line(-1,0){1}} \put(15.6,5.4 ){\line(0,-1){0.8}}
\put(14.8,4.8){G$_3^{(2)}$}


\put(1,4.25){\line(0,-1){0.5}}
\put(2.45,5){\vector(1,0){12.1}}

 \put(1,3.75){\vector(1,0){13.6}}

\put(14.6,3.4){\line(1,0){1}} \put(14.6,3.4 ){\line(0,1){0.8}}
\put(15.6,4.2){\line(-1,0){1}} \put(15.6,4.2 ){\line(0,-1){0.8}}
\put(14.9,3.6){G$_4$}


\put(2.5,11){\line(1,0){1.7}} \put(2.5,11){\line(0,1){0.8}}
\put(2.5,11.8){\line(1,0){1.7}}

\put(4.2,11.8){\line(1,-1){0.4}}
 \put(4.2,11){\line(1,1){0.4}}
 \put(2.6,11.25){$Z\! \equiv\! Z(\Sigma)$}

\put(4.6,11.4){\line(1,0){0.25}} \put(4.85,11.4){\line(0,1){0.7}}


\put(2.5,14.2){\line(1,0){1.7}} \put(2.5,14.2){\line(0,1){0.8}}
\put(2.5,15){\line(1,0){1.7}}

\put(4.2,15){\line(1,-1){0.4}}
 \put(4.2,14.2){\line(1,1){0.4}}
 \put(2.6,14.45){$Z\! \equiv\! Z(S)$}

\put(4.6,14.6){\line(1,0){0.2}} \put(4.8,14.6){\line(0,-1){0.7}}


\put(5.5,10.35){\line(1,0){1.7}} \put(5.5,10.35){\line(0,1){0.8}}
\put(5.5,11.15){\line(1,0){1.7}}

\put(7.2,11.15){\line(1,-1){0.4}}
 \put(7.2,10.35){\line(1,1){0.4}}
 \put(5.6,10.6){$Z\! \equiv\! Z(A)$}

\put(7.6,10.75){\line(1,0){0.4}} \put(8,10.75){\line(0,-1){0.75}}

\put(5.5,6.8){\line(1,0){1.7}} \put(5.5,6.8){\line(0,1){0.8}}
\put(5.5,7.6){\line(1,0){1.7}}

\put(7.2,7.6){\line(1,-1){0.4}}
 \put(7.2,6.8){\line(1,1){0.4}}
 \put(5.6,7.05){$Z\! \equiv\! Z(Y)$}

\put(7.6,7.2){\line(1,0){0.5}} \put(8.1,7.2){\line(0,1){0.8}}


\put(3.2,15.9){yes} \put(1.2,14.6){no}

\put(2.5,13.5){no} \put(2.5,12.3){no}

\put(2.7,9.6){no} \put(2.5,12.3){no} \put(2.7,8.2){no}
 \put(2.7,5.2){no}

 \put(1.2,12.9){yes}

\put(1.2,11.05){yes}

\put(1.2,8.9){yes}

 \put(1.2,6.5){yes}

\put(1.2,3.95){yes}

\put(8.5,13.7){yes}

\put(9.8,13.1){no} \put(13,13.1){yes}

\put(11.7,12.2){no} \put(11.7,10.8){no}

\put(13,11.8){yes}

\put(11.7,8.2){no} \put(11.7,6.8){no} \put(13,9.2){yes}

\put(13,7.8){yes}
\end{picture}}

\vspace{-30mm}
\caption{This flow diagram distinguishes the dimension of the groups of isometries of a three-dimensional Riemannian metric.}
\label{figure-1}
\end{figure} 


This flow diagram uses as initial input data the metric $g$, the Ricci tensor $R$, the algebraic Ricci concomitants $S$, $r$, $s$ and $t$ defined in (\ref{Srst}), and the first-order Ricci concomitants $\Sigma$, defined in (\ref{Sigma}), and $\dif \alpha$, defined in (\ref{A}). In subsequent steps we need the connection tensor $Z(T)$ given in (\ref{Z(T)}), where $T$ is a tensor that depends on the different cases: $Z(S)$ is of first order in the Ricci tensor, and $Z(\Sigma)$, $Z(A)$ and $Z(Y)$ are of second order. Concomitants $A$ and $Y$ are given in (\ref{A}) and (\ref{Y}), respectively. From $Z(T)$ we can define the tensors $C(Z)$ given in (\ref{C(Z)}) (first-order in $Z$), and $D(Z)$ given in (\ref{D(Z)}) (second-order in $Z$). Conditions H1 and H2 are specified in (\ref{H2b}) and (\ref{H1b}), respectively. They use the algebraic concomitants $I, J, L, M, N$ of $C(Z)$ and $D(Z)$ given in (\ref{IJ}) and (\ref{LMN}). The end horizontal arrows lead to the different G$_r$. G$_3^{(3)}$ and G$_3^{(2)}$ denote the action of a three-dimensional group on three dimensional and on two-dimensional orbits, respectively. The two end vertical arrows lead to non-existence of isometries.

\section{Labeling the several G$_3^{(2)}$ groups and the commutative G$_2$ group}
\label{sec-G3(2)-G2}

In \cite{FS-G3} we have analyzed the groups of isometries in transitive action on the whole space, and we have also distinguished the three different groups G$_6$, the three different groups G$_4$ and the ten Bianchi-Behr types G$_3$. Now we characterize the three different G$_3^{(2)}$, and when a group G$_2$ is commutative.

When a three-dimensional group of isometries acts on two-dimensional orbits, these orbits are of constant curvature $k$. And the group is SO(3) when $k=+1$, SO(2,1) when $k=-1$, and E(2) when $k=0$. Bona and Coll \cite{bonacoll2} gave the following invariant expression for this curvature, $k= {\rm sign} \left( \frac{\theta^2}{4} + \beta - \frac{\alpha}{2} \right)$, where $\theta$ is the expansion of the simple eigenvector of the Ricci tensor and $\alpha$ and $\beta$ are, respectively, the simple and the double Ricci eigenvalues. One can easily obtain these invariants in terms of explicit Ricci concomitants already used in this paper, and we can state:
\begin{proposition}
The three-dimensional Riemannian spaces that admit a three-dimensional group of isometries acting on two-dimensional orbits are characterized by the algorithm in figure {\em \ref{figure-1}}. The group is {\em SO(3)} when $k=+1$, {\em SO(2,1)} when $k=-1$, and {\em E(2)} when $k=0$, with $k$ defined as:
\be
k = {\rm sign}\left\{\frac14 (\nabla \cdot h)^2 +  \frac{1}{6} r - 2 \frac{t}{s}\right\} \, ,
\ee
where $s$ and $t$ are given in {\em (\ref{Srst})}, and $h$ is given in {\em (\ref{Sigma})}.
\end{proposition} 

When the Riemannian space admits a group G$_2$ of isometries, the isotropy group is trivial and a Ricci-frame exists. If $Z$ is its associated connection tensor, then the covariant derivative of any Killing vector $\xi$ can be obtained as \cite{FS-G3}: $\nabla \xi = *(\xi \cdot Z)$. Consequently, if $\xi_1$, $\xi_2$ are two independent Killing vectors, we obtain:
\be
[\xi_1 , \xi_2] = \xi_1 \cdot *(\xi_2 \cdot Z) - \xi_2 \cdot *(\xi_1 \cdot Z)  =  *\left((\xi_1 \wedge \xi_2) \cdot Z\right) \, .
\ee
Thus, the group G$_2$ is commutative if, and only if, $*((\xi_1 \wedge \xi_2) \cdot Z)=0$. 
On the other hand, we know that the first index of the tensor $C(Z)$ defines a sole direction when a maximal group G$_2$ of isometries exists, and this direction is orthogonal to the orbits, that is, it is given by $*(\xi_1 \wedge \xi_2)$. Therefore, the commutative condition can be written as $\eta^{i j k} C_{k m n} {Z_{j}}^p \, \eta_{p i l } =0$. Then, a straightforward calculations leads to:
\begin{proposition}
The three-dimensional Riemannian spaces that admit a two-dimensional group of isometries are characterized by the algorithm in figure {\em \ref{figure-1}}. The group is commutative if, and only if,
\be
Z \cdot C - (\tr Z)\, C = 0 \, ,
\ee
where $Z=Z(T)$ is given in {\em (\ref{Z(T)})}, with $T=S, \Sigma, A, Y$, depending on the different cases, and $C=C(Z)$ is given in {\em (\ref{C(Z)})}. \end{proposition} 
%


\ack This work has been supported by the Spanish Ministerio de Ciencia, Innovaci\'on y Universidades and the Fondo Europeo de Desarrollo Regional, Projects PID2019-109753GB-C21 and PID2019-109753GB-C22, the Generalitat Valenciana Project AICO/2020/125 and the University of Valencia Special Action Project  UV-INVAE19-1197312.


\end{document}